\documentclass[twocolumn,showpacs,amsmath,amssymb,pre,10pt]{revtex4}
\usepackage{amssymb}
\usepackage[final]{graphicx}

\usepackage{epstopdf}

\begin{document}

\title{Effects of Annealing Conditions on the Microstructure and Magnetic Properties of the Perovskite Manganite, $\text{La}_{0.75}\text{Sr}_{0.25}\text{MnO}_{3}$ }
\author{D.O.J. Green and K-U. Neumann}

\affiliation{Department of Physics, Loughborough University,
Loughborough, LE11 3TU, United Kingdom\\}

\begin{abstract}
The effects of annealing conditions upon the microstructure and the magnetic properties of the colossal magnetoresistive manganite $\text{La}_{0.75}\text{Sr}_{0.25}\text{MnO}_{3}$ have been investigated. Increasing the annealing temperature and time of annealing is seen to increase the size of crystallites within the samples. The spontaneous magnetic moment per formula unit and the Curie temperature, as obtained from Arrott plot analysis, are observed to depend upon the average size of crystallites.
\end{abstract}

\pacs{61.72.-y, 75.47.Lx, 75.47.-m}

\maketitle 
\section{Introduction}

The existence of a ferromagnetic-paramagnetic and metal-insulator transition in lanthanum manganites was established in
the early 1950's \cite{jon, vans} and has been extensively studied thereafter. The magnetic transition which is associated with unusual transport properties, including large negative magnetoresistance, is observed in a family of doped manganites of perovskite structure, with the chemical formula $\text{RE}_{1-x}\text{A}_{x}\text{MnO}_{3}$, where RE is a rare earth (La, Pr, Nd, Sm...), and A is a divalent
metal (Ca, Sr, Ba...).
The resurgence of interest in these systems is related to the demonstration of very large
negative magnetoresistance in thin films \cite{vonH, jin}, termed colossal magnetoresistance (CMR).

The metal insulator transition in manganites has traditionally been attributed to the double exchange mechanism, which results in a varying bandwidth of electrons in the $\text{Mn}^{3+}$ d shell as a function of temperature and doping level \cite{DE, DEpapers, eff DE}. However, more recently it has been realized \cite{millis, millis2} that the effective carrier-spin exchange interaction of the double-exchange model is too weak to lead to a significant reduction of the electron bandwidth. A giant isotope effect \cite{zhao, Oiso}, the sign anomaly of the Hall effect and the Arrhenius behaviour of the drift and hall mobilities, and the fact that polaron hopping satisfactorily accounts for resistivity in the paramagnetic phase \cite{Heff}, suggest that the charge carriers in perovskite manganites are of polaronic character.
 Low-temperature optical \cite{optical, optical2, optical3}, electron energy-loss (EELS) \cite{EELS}, photoemission \cite{dessau, chuang}, and thermoelectric \cite{zhou} measurements showed that the ferromagnetic phase of manganites is not a conventional metal and confirmed that manganites were, in fact, charge-transfer doped insulators having oxygen p holes as current carriers as opposed to d $\text{Mn}^{3+}$ electrons proposed in the double exchange theory. These observations led to a novel theory of the ferromagnetic-paramagnetic phase transition driven by non-degenerate polarons in doped charge-transfer magnetic insulators, the so-called current-carrier density collapse (CCDC) theory \cite{AS}.
 The essence of the CCDC theory is that in the paramagnetic phase a large fraction of the polaronic carriers are bound as immobile bipolarons, that break-up into single polarons on thermal excitation. In the ferromagnetic phase the exchange interaction between the localized $\text{Mn}^{3+}$ electrons and the oxygen p-holes, causes a break-up of immobile bipolarons into single polarons. Within the CCDC theory an applied magnetic field has the effect of breaking up bipolaronic pairs, giving rise to an increased number of charge carriers and hence, a reduction in resistivity.
 
 A combination of low-energy electron diffraction and angle-resolved photoemission spectroscopy on some semiconducting interfaces has provided unambiguous evidence of a bipolaronic insulating state \cite{bipolaronic insulator, prb bipolaronic insulator}. There is also evidence to support the existence of immobile spin-singlet bipolarons in barium titanate perovskites \cite{bipolaron in perovskites, singlet bipolaron in perovskites} which are of a very similar structure to the CMR perovskite manganites. Furthermore x-ray and neutron scattering measurements directly demonstrate the existence of short-range correlations of polarons in the paramagnetic phase of colossal magnetoresistive perovskite manganites \cite{charge melting and polaron collapse, short range polaron correlations, correlated polarons, charge correlations}. The polaron correlations are shown to have correlation lengths of approximately 10$\text{\AA} $ \cite{charge correlations}. These short-range polaron correlations are shown to grow with decreasing temperature, but disappear abruptly at the ferromagnetic transition \cite{charge melting and polaron collapse, short range polaron correlations, charge correlations}. It is established that the temperature dependence of the polaron correlations is intimately related to the transport properties of the manganites \cite{short range polaron correlations, charge correlations}. Also the polaron correlations collapse under an applied magnetic field \cite{charge melting and polaron collapse, charge correlations}. The polaron correlations are interpreted as bipolarons by Nelson et al. \cite{correlated polarons}. The experimental observations of polaron correlations in the paramagnetic phase of perovskite manganites are fully consistent with the formation of bipolarons within the CCDC theory.

Due to the close relationship between the magnetic and transport properties within the perovskite manganites it is important to understand the factors affecting the magnetic characteristics within these materials.
In the present paper we investigate the effects of annealing temperature and annealing time upon the microstructure and magnetic properties of the perovskite $\text{La}_{0.75}\text{Sr}_{0.25}\text{MnO}_{3}$ prepared by a standard solid state reaction.

\section{Sample Preparation and Structural Characterization of $\text{La}_{0.75}\text{Sr}_{0.25}\text{MnO}_{3}$}
\label{sample prep}

Samples of $\text{La}_{0.75}\text{Sr}_{0.25}\text{MnO}_{3}$ have been prepared using a standard solid state reaction method, from powders of $\text{La}_{2}\text{O}_{3}$, $\text{SrCO}_{3}$ and $\text{MnO}_{2}$. These powders were checked for water content by examining x-ray diffraction profiles before and after heating. The x-ray examination showed that the initial constituents contained little or no water. However, to eliminate any chance of water content affecting sample preparation, heated constituent chemicals were used in the production of $\text{La}_{0.75}\text{Sr}_{0.25}\text{MnO}_{3}$. The prescribed masses of the constituent chemicals to produce 15g of our $\text{La}_{0.75}\text{Sr}_{0.25}\text{MnO}_{3}$ were thoroughly mixed together and then pressed into pellets. 

\begin{figure}[h]
	\centering
		\includegraphics[scale=0.78]{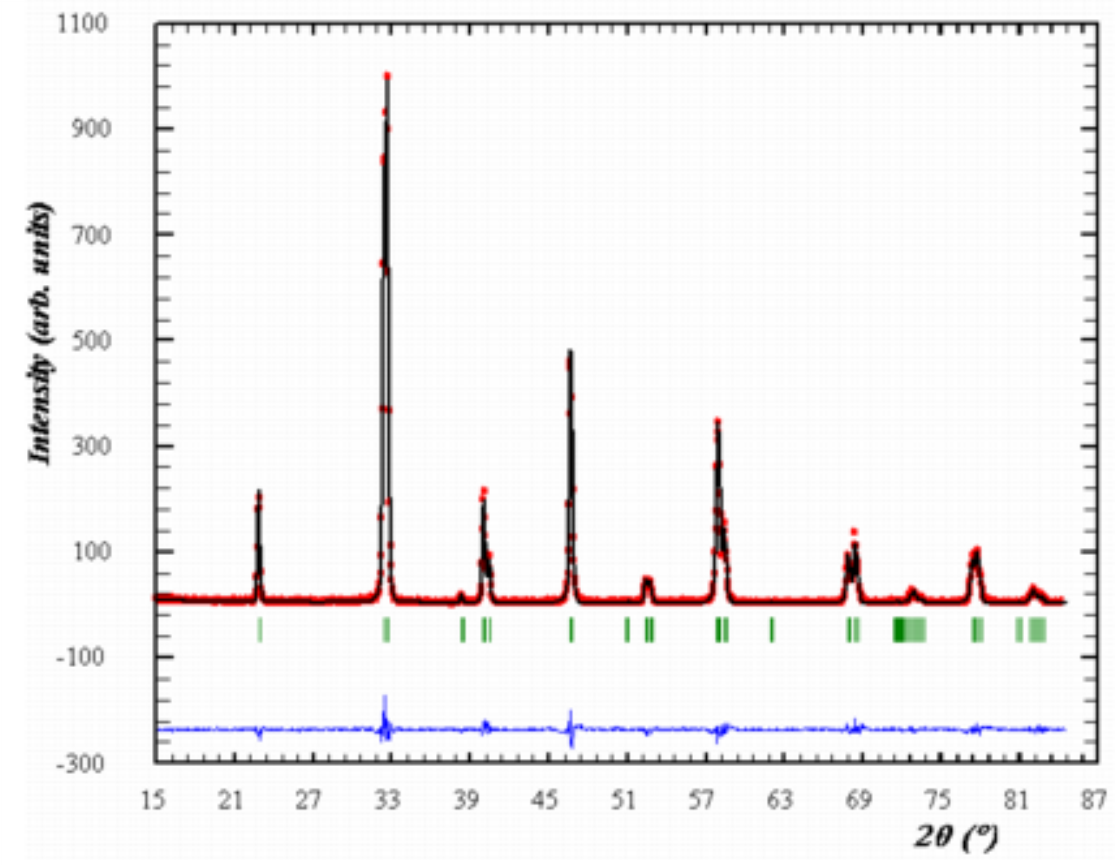}
		\caption{The x-ray diffraction pattern produced after the final 24 hour, $1100^\circ\text{C}$ sintering process, as described in the text, with a Fullprof Rietveld refinement \cite{fullprof}.}	
		\label{fulproffit}	
\end{figure}

These pellets were placed into a ceramic heating vessel and reacted/sintered, initially at $400^\circ\text{C}$ for three days. The pellets were then ground into a fine powder, mixed thoroughly and pressed into pellets again. The process was repeated, the temperature of each heating/sintering process was increased in $200^\circ\text{C}$ increments to: $600^\circ\text{C}$, $800^\circ\text{C}$, $1000^\circ\text{C}$ and finally $1100^\circ\text{C}$. The powder x-ray diffraction profile produced after the final $1100^\circ\text{C}$ heating/sintering process is presented in fig (\ref{fulproffit}).

Rietveld refinement in the FullProf program \cite{fullprof} suggests the sample of $\text{La}_{0.75}\text{Sr}_{0.25}\text{MnO}_{3}$ produced has the rhombohedral perovskite structure with space group $R\overline{3}c$, with lattice parameters: $\text{a}=\text{b}= 5.5212(7)\text{\AA}$, $\text{c}=13.379080(0)\text{\AA}$, with $\alpha=\beta= 90^\circ$ and $\gamma= 120^\circ$, giving a cell volume of  353.939( 0.044) $\text{\AA}^{3}$.

The $\text{La}_{0.75}\text{Sr}_{0.25}\text{MnO}_{3}$ sample was pressed into 3 separate pellets, each pellet was subjected to further, different annealing processes to produce 3 samples each with different microstructural properties. 
 The different annealing conditions used were, $900^\circ\text{C}$ for 24 hours, $1200^\circ\text{C}$ for 24 hours and $1200^\circ\text{C}$ for 84 hours. The samples produced were then analysed using powder x-ray diffraction. Structural and microstructural refinements of the resulting powder x-ray diffraction patterns was undertaken using the FullProf Rietveld refinement program \cite{fullprof}. 

Crystallite size and anisotropic strain are related to the broadening of the x-ray diffraction profile lines. The resolution of the diffractometer also contributes to this broadening, and as such must be subtracted. A silicon standard sample was used to find the instrumental resolution of the Siemens D5000 diffractometer, as silicon displays negligible line broadening. The anisotropic size broadening effects, related to the coherence volume of diffraction, are simulated in the FullProf program \cite{fullprof} using spherical harmonics. The spherical harmonics are referred to in a Cartesian frame of reference where the z-axis is always along the principal symmetry axis. In the case of a trigonal system the hexagonal setting is used and ``c'' is along the z-axis. The order of the harmonics is arbitrarily limited as it is the case for strain broadening. A quadratic form of the anisotropic strain effects is also employed \cite{juan rodriguez}.

\begin{figure}[h]
	\centering
		\includegraphics[scale=0.85]{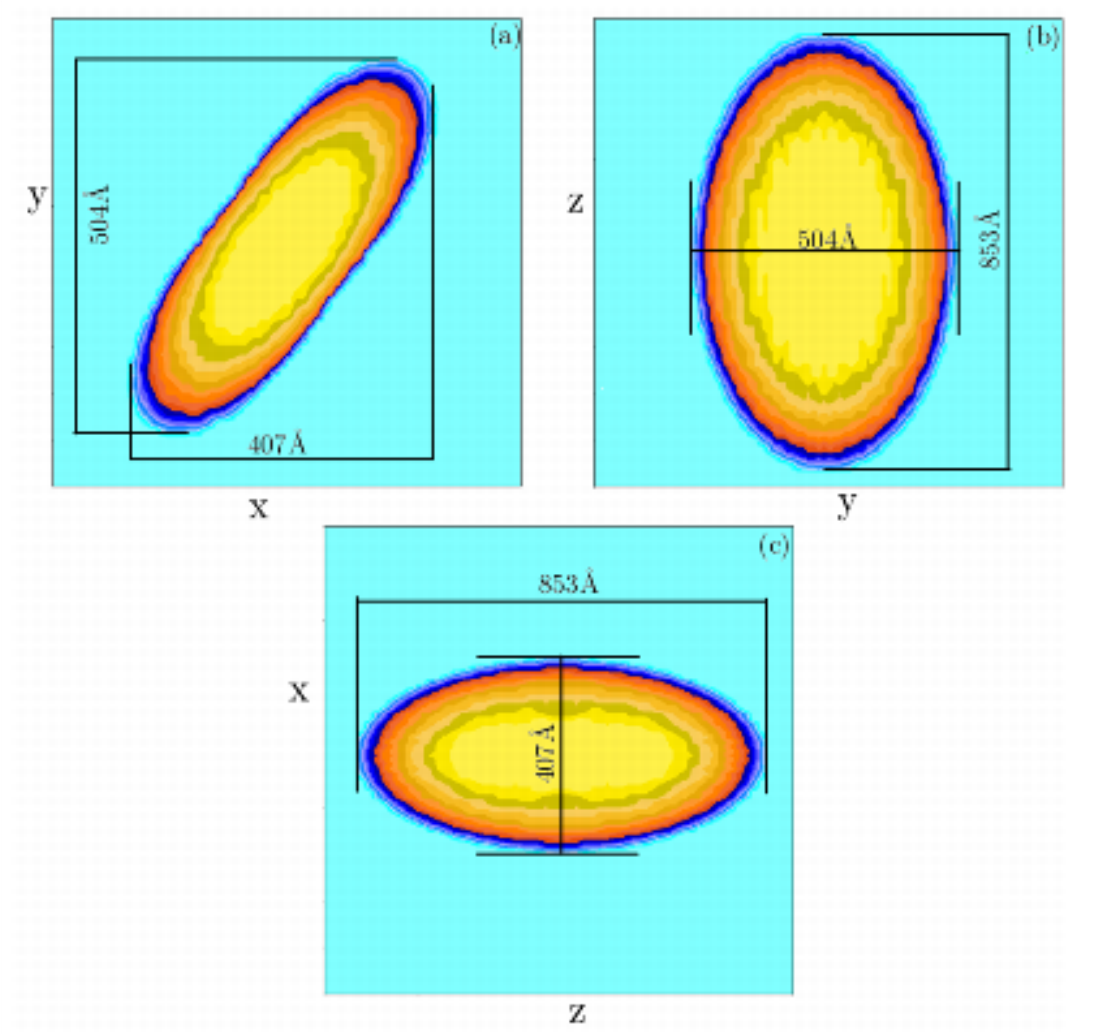}
		\caption{Projections of the crystallite shape for the sample produced after annealing at $1200^\circ\text{C}$ for 84 hours in the crystallographic planes: (a) (001), (b) (100), (c) (010). Images produced using the Gfourier program \cite{gfourier} in conjunction with the FullProf program \cite{fullprof}.   }	
		\label{crystallite shape}	
\end{figure}

The structural and microstructural refinements, reveal that the different annealing conditions have little effect upon the crystallographic structure of $\text{La}_{0.75}\text{Sr}_{0.25}\text{MnO}_{3}$. However, the annealing conditions play a major role in the size of crystallites while the anisotropic lattice strain varies slightly. Visualisations of the average crystallite shape obtained by using the Gfourier program \cite{gfourier} show that the crystallite shape is very similar for all samples regardless of annealing conditions. All samples are composed of ``disk'' shaped crystallites, see fig(\ref{crystallite shape}) for an example. The FullProf program \cite{fullprof} calculates the apparent crystallite size for each crystallographic direction. The average of these apparent crystallite sizes for each sample are shown in table (\ref{magnetic parameters}).

\section{Magnetic Properties of $\text{La}_{0.75}\text{Sr}_{0.25}\text{MnO}_{3}$}
\begin{figure}[h]
	\centering
		\includegraphics[scale=1.05]{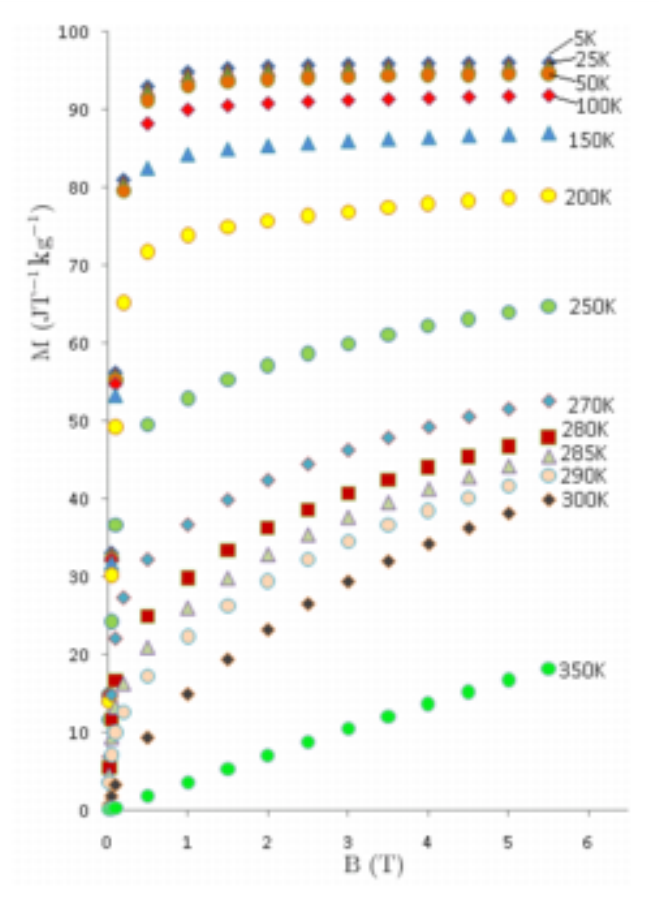}
		\caption{Isothermal magnetization data of $\text{La}_{0.75}\text{Sr}_{0.25}\text{MnO}_{3}$ annealed at $1200^\circ\text{C}$ for 84 hours.}	
		\label{isotherms}	
\end{figure}

Isothermal magnetization data for samples of $\text{La}_{0.75}\text{Sr}_{0.25}\text{MnO}_{3}$ with different annealing conditions has been obtained from measurements performed on a Superconducting Quantum Interference Device (SQUID). The particular model was a Quantum Design Magnetic Property Measurement System (MPMS). The isothermal magnetization data of $\text{La}_{0.75}\text{Sr}_{0.25}\text{MnO}_{3}$ annealed at $1200^\circ\text{C}$ for 84 hours is presented in fig (\ref{isotherms}).

The investigation of magnetic properties with the use of Arrott plots is a useful tool in the
analysis and description of magnetic materials. The basis of Arrott plots originates from a Landau description of the magnetization, in which the free energy has the form \cite{klaus book};

\begin{equation}
F=F_{0}+ \frac{AM^{2}}{2}+\frac{CM^{4}}{4}-\mu_{0}MB
\label{F}
\end{equation}

Here A and C are coefficients in a series expansion of the free energy F in powers of the magnetic moment M. 
The external magnetic field is given by $\mu_{0}B$, and $F_{0}$ is comprised of all contributions to the free energy that are independent of the magnetic moment M.
 The C coefficient is taken to be temperature independent, whereas the  A coefficient is dependent upon temperature  and is related to the magnetic susceptibility $\chi$(T) as; $\chi$(T)= 1/A(T).
 Minimizing eq (\ref{F}) we arrive at:

  \begin{eqnarray}
\frac{\partial F_{0}}{\partial M}&=& AM + CM^{3}-\mu_{0}B=0 \nonumber \\
&\Rightarrow& \mu_{0}B=AM + CM^{3}      \nonumber \\
   \label{Fminimizing}
\end{eqnarray}

Arrott \cite{arrot} noticed that eq (\ref{Fminimizing}) may be written as;

\begin{eqnarray}
M^{2}=\frac{1}{C}\frac{\mu_{0}B}{M}-\frac{A}{C}
\label{arroteq}
\end{eqnarray}

Arrott plots are obtained by plotting the values of M(H) and B as a function of M$^{2}$ versus $\mu_{0}$B/M. According to eq (\ref{arroteq}) this should result in a straight line. There are, however, some limitations to this approach. For $\text{T}\approx \text{T}_{\text{C}}$ critical fluctuations become important and one should expect deviations from the straight line Arrott plot \cite{borovik}. Also at low values of the applied field, deviations from straight line behaviour occur even for a chemically homogeneous sample. These deviations may arise due to inhomogeneities, such as the magnetic domain structure. These inhomogeneities influence the magnetic moment and result in the low field curvature of the Arrott plots.
At temperatures below the Curie temperature, $\text{T}_{\text{C}}$, the coefficient A is negative, and at temperatures greater than  T$_{\text{C}}$ the coefficient A is positive \cite{klaus book}. $\text{T}_{\text{C}}$ is therefore defined as the temperature for which the straight line Arrott plot intercepts the origin.

\begin{figure}[]
	\centering
		\includegraphics[scale=0.62]{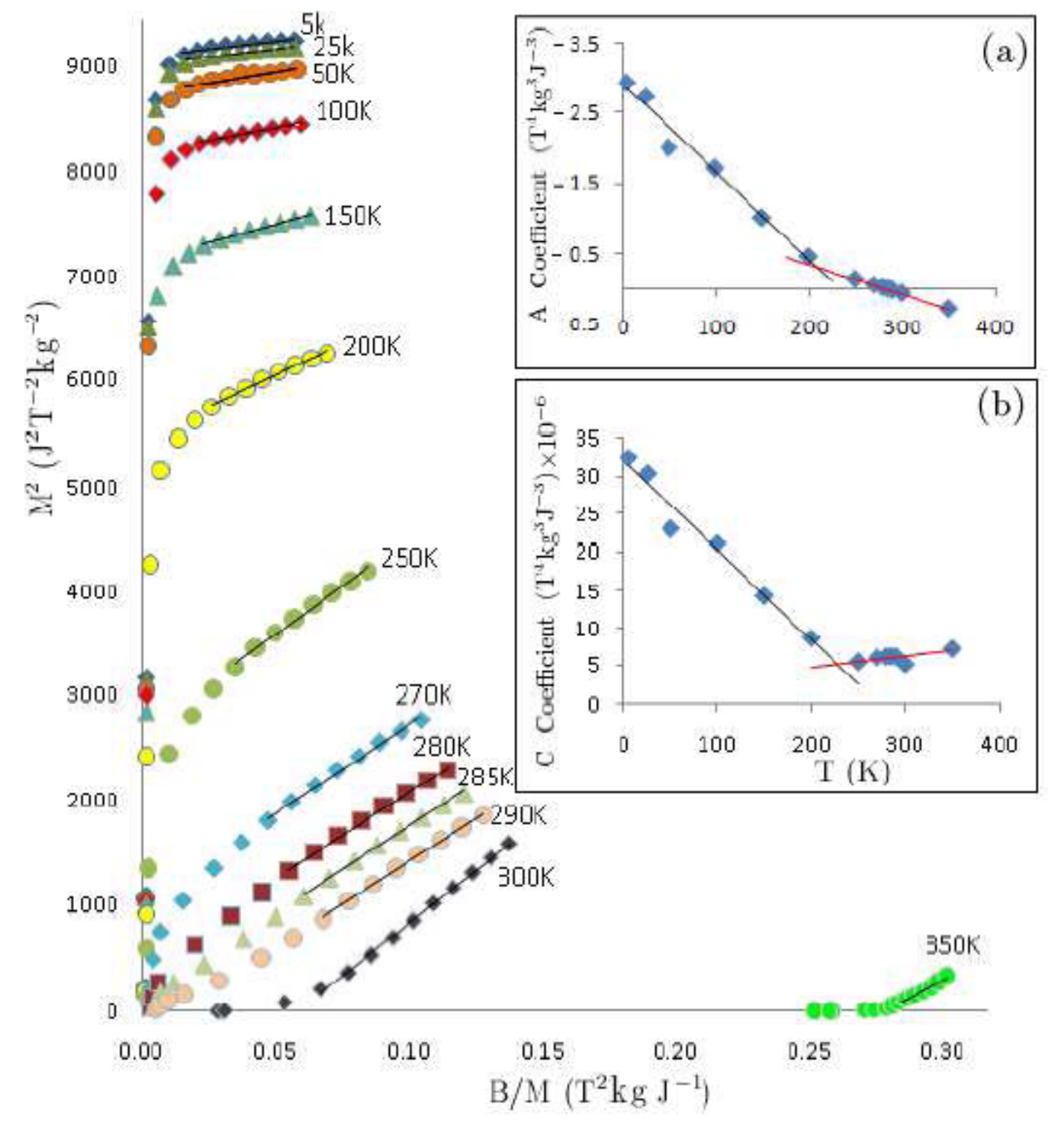}
		\caption{Arrot plots of $\text{La}_{0.75}\text{Sr}_{0.25}\text{MnO}_{3}$ annealed at $1200^\circ\text{C}$ for 84 hours, constructed from the isothermal magnetization data presented in fig (\ref{isotherms}), including linear fits to the high field data. Insert (a) shows the temperature dependence of the A coefficient, and insert (b) shows the temperature dependence of the C coefficient. The solid lines in the inserts are guides to the eye.}
\label{arrott}	
\end{figure}

Arrott plots have been produced for all samples of $\text{La}_{0.75}\text{Sr}_{0.25}\text{MnO}_{3}$ with different annealing conditions.  The Arrott plots constructed for the sample of $\text{La}_{0.75}\text{Sr}_{0.25}\text{MnO}_{3}$ annealed at $1200^\circ\text{C}$ for 84 hours, along with the temperature dependence of the A and C coefficients, are presented in fig(\ref{arrott}).

By plotting the y intercept values from the Arrott plots as a function of temperature, the spontaneous ferromagnetic moment is obtained at zero applied field, and the value of $\text{T}_{\text{C}}$ is interpolated.

The coefficients A and C are plotted in fig (\ref{arrott}) as a function of temperature. Both the A and C coefficients are seen to decrease steadily with increasing temperature until a change in gradient is observed between 200-250 K. Such a change in the temperature dependence of the A and C coefficients is indicative of a change in the magnetic characteristics of the sample, at approximately this temperature interval. This change in the magnetic characteristics is separate from the Curie temperature $\text{T}_{\text{C}}$ which is found as 285.6 K. All samples exhibit a weak temperature dependent C coefficient. 

Values of the spontaneous magnetic moment per formula unit at 0 K and the Curie temperature for each sample are given in table (\ref{magnetic parameters}).

\begin{table}[h!]
\begin{tabular}{|l | c |c|c| }
\hline
Annealing Temperature ($^{0}\text{C}$)&900& 1200&1200 \\ 
 \hline
Annealing Time (hours) &24&24&84\\ 
 \hline
  Average apparent Crystallite Size (\AA)& 519.17 &  706.95 &  797.05 \\   
 \hline
   Spontaneous Magnetic Moment/f.u ($\mu_{B}$)& 3.04&3.70&3.91 \\   
 \hline
   Curie Temperature (K) &336.7  &349.5 & 285.6\\   
 \hline
 \end{tabular}
 \caption{A table summarizing the microstuctural and magnetic properties of $\text{La}_{0.75}\text{Sr}_{0.25}\text{MnO}_{3}$ as a function of annealing conditions. The average crystallite size corresponds to a volume conserving sphere of the coherent volume of diffraction.}
 \label{magnetic parameters}
\end{table}

\begin{figure}[h]
	\centering
		\includegraphics[scale=0.7]{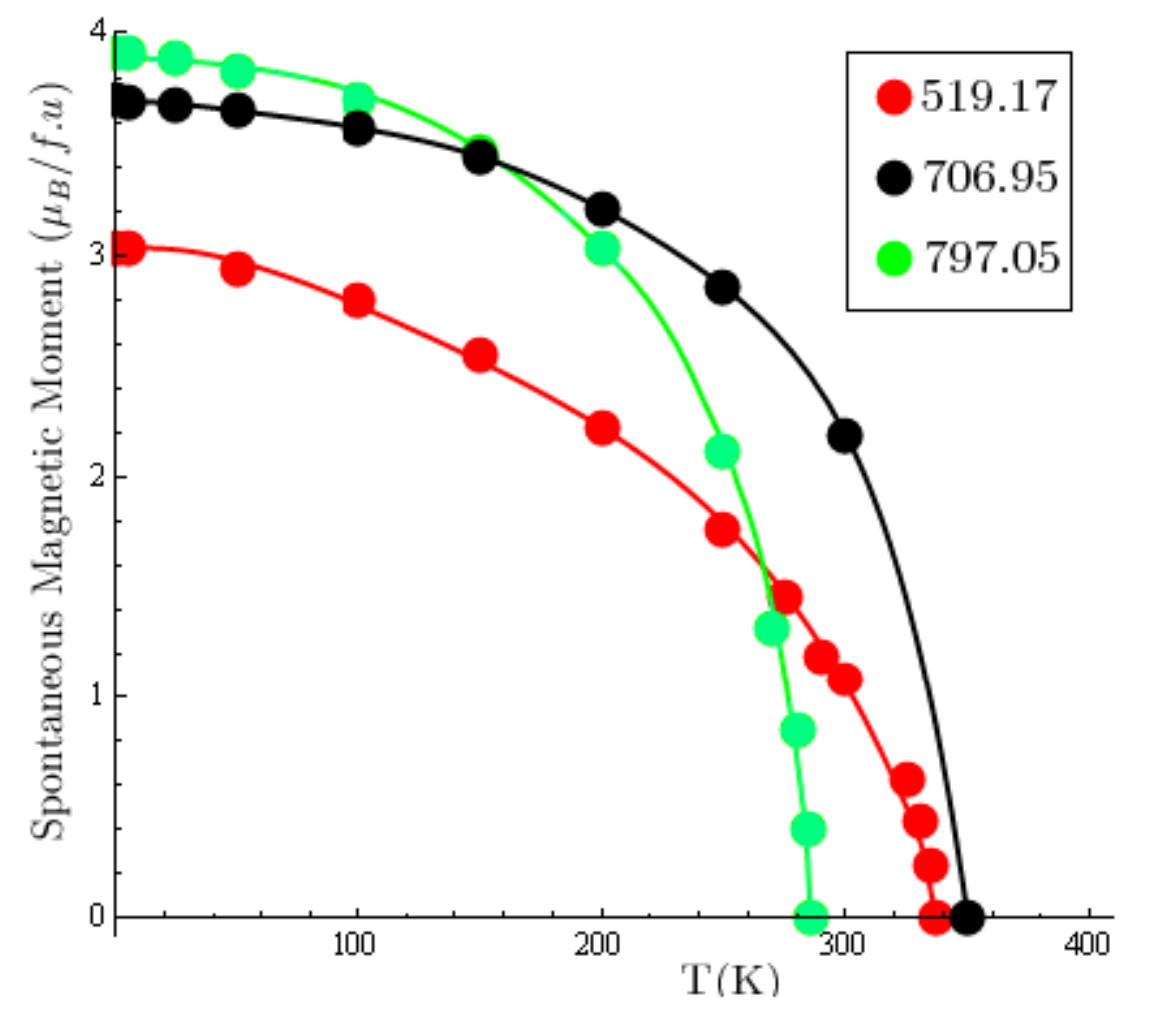}
		\caption{Plots of the spontaneous moment of $\text{La}_{0.75}\text{Sr}_{0.25}\text{MnO}_{3}$ as a function of temperature, for samples composed of different sized crystallites given in Angstroms. The solid lines are guides to the eye.}
		\label{moments}	
\end{figure}

The spontaneous moment as a function of temperature is plotted in fig (\ref{moments}). The transition becomes much sharper with increasing crystallite size. The value of the spontaneous magnetic moment also increases as the average crystallite size is increased by altering the annealing conditions. The sample of $\text{La}_{0.75}\text{Sr}_{0.25}\text{MnO}_{3}$ with an average apparent crystallite size of 519.17 $\text{\AA}$ has a Curie temperature of 336.7 K, as the average apparent crystallite size is increased to 706.95 $\text{\AA}$ $\text{T}_{\text{C}}$ increases slightly to 349.5 K, further increase of the average apparent crystallite size to 797.05 $\text{\AA}$  actually decreases $\text{T}_{\text{C}}$ to 285.6 K.

\section{Discussion of Results}

\begin{figure}[h!]
	\centering
		\includegraphics[scale=0.6]{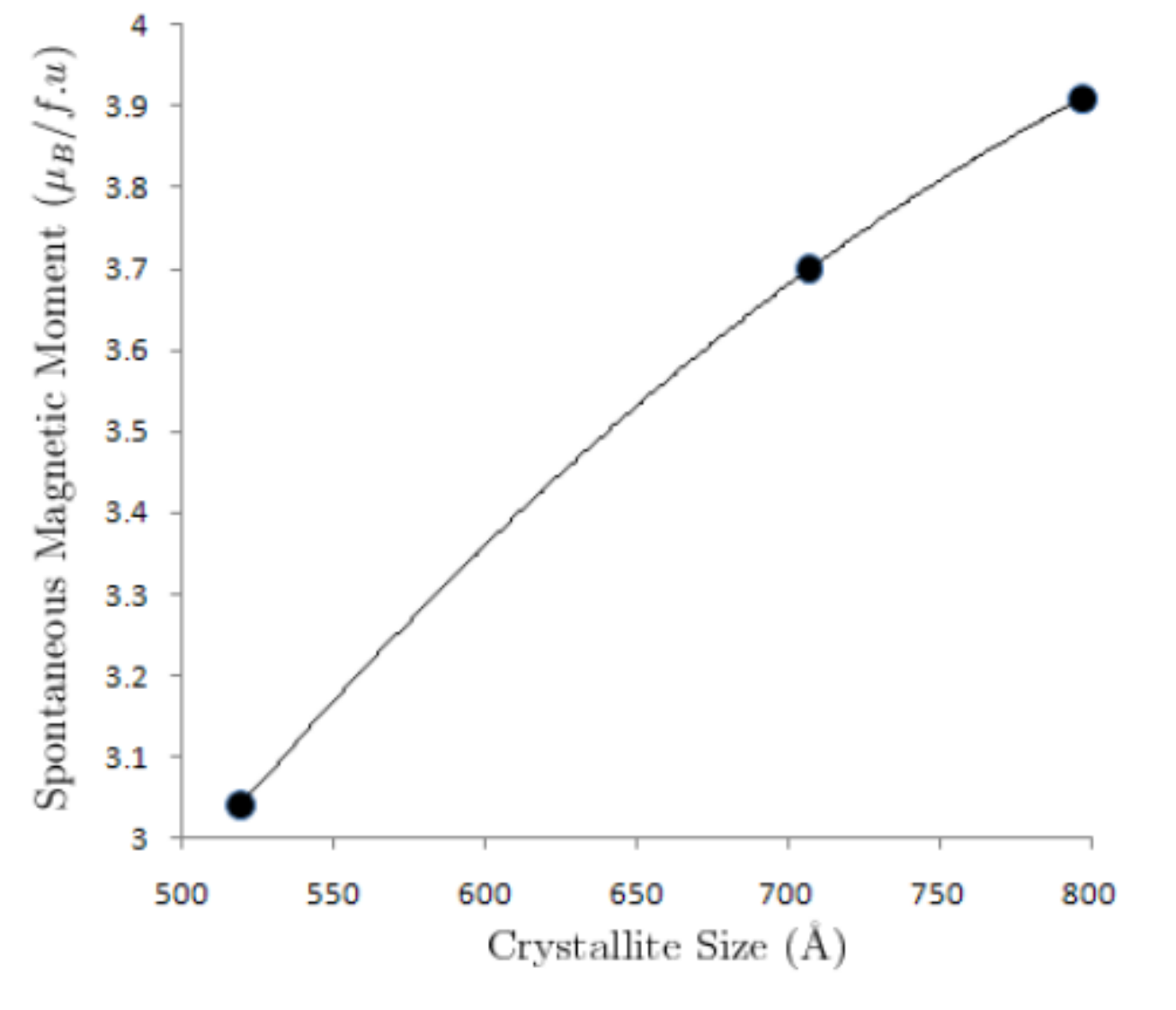}
	\caption{A plot of the spontaneous moments as a function of crystallite size as determined by micro-structural refinement using the FullProf Rietveld refinement program. The solid line is a guide to the eye.}
	\label{moment size}
\end{figure}

One can see that the spontaneous moment at 0 K, increases dramatically as the crystallite size is increased, as shown in fig (\ref{moment size}). Increasing crystallite size will reduce the amount of amorphous sample found within the inter-grain boundaries, which will most likely not display a spontaneous magnetic moment. Hence reducing the density of grain boundaries by increasing the temperature and duration of the annealing process should lead to an increase in the observed spontaneous moment per formula unit, as evidenced by the results presented here. 
 A noticeable result is that as the spontaneous moment is increased from 3.04 $\mu_{B}$ to 3.91 $\mu_{B}$ the Curie temperature actually decreases from 336.7 K to 285.6 K. Within a mean field model one would expect an increase in the magnetic moment to lead to an increase in $\text{T}_{\text{C}}$. It therefore appears that the mean field theory is insufficient to describe the magnetic properties of bulk samples of $\text{La}_{0.75}\text{Sr}_{0.25}\text{MnO}_{3}$.

The coherently diffracting regions of the $\text{La}_{0.75}\text{Sr}_{0.25}\text{MnO}_{3}$ samples are probed using powder x ray diffraction techniques. As the size of the coherently diffracting regions is decreased finite size effects will play an ever more important role in determining the properties of the $\text{La}_{0.75}\text{Sr}_{0.25}\text{MnO}_{3}$ samples, such as the band structure within the crystallites and the coupling between crystallites.

The broadening of the transition and the curious result that the sample with the lowest Curie temperature exhibits the largest spontaneous moment, are allowed for within the theoretical framework of the current carrier density collapse (CCDC) theory \cite{AS}. Within this theory the transition temperature is directly related to the bipolaron binding energy. FullProf Rietveld refinement has shown that the different annealing conditions have little effect upon the crystallographic structure of $\text{La}_{0.75}\text{Sr}_{0.25}\text{MnO}_{3}$. However, the refinement has shown that the average crystallite size is dramatically altered due to different annealing conditions.  Each crystallite may exhibit locally different bipolaron binding energies. In samples with smaller crystallite size there is most likely a greater range of bipolaron binding energies present.

 Alexandrov et al. \cite{coex}, have considered a model of colossal magnetoresistive manganites in which paramagnetic and ferromagnetic regions coexist due to a distribution of local bipolaron binding energies across the sample, caused by disorder. The coexistence of ferromagnetic and paramagnetic regions within this model causes a broadening of the transition about the mean value of $\text{T}_{\text{C}}$. Therefore reducing the level of disorder within this model could reduce the maximum temperature at which a spontaneous magnetic moment is observed. The value of $\text{T}_{\text{C}}$ determined by Arrott plot analysis gives the maximum temperature at which the bulk sample exhibits a spontaneous moment. A decrease in this value of $\text{T}_{\text{C}}$ in more ordered samples that exhibit a larger spontaneous moment, fits within the phase coexistence model of colossal magnetoresistive manganites \cite{coex}. The anisotropic lattice strain could also play a role in the broadening of the transition, however this was seen to vary only slightly with different annealing conditions. 

S. Yang and C.T. Lin \cite{daas} have produced samples of $\text{La}_{0.83}\text{Sr}_{0.17}\text{MnO}_{3}$ by deposition from aqueous acetate solution. In the samples of $\text{La}_{0.83}\text{Sr}_{0.17}\text{MnO}_{3}$, the crystallite size was increased by annealing at higher temperatures. This led to an increase in the saturation magnetic moment. Also the most ordered sample i.e. the sample with the largest crystallite size exhibited the lowest Curie temperature. We have observed similar results in bulk samples of $\text{La}_{0.75}\text{Sr}_{0.25}\text{MnO}_{3}$ prepared using a totally different method. We have also observed an increase in spontaneous magnetic moment with increasing crystallite size, and that the ordered sample, $\text{La}_{0.75}\text{Sr}_{0.25}\text{MnO}_{3}$ annealed at $1200^{\circ}\text{C}$ for 84 hours, displays the lowest Curie temperature. Our results, however, show a much greater variation in T$_{\text{c}}$ with crystallite size.

\end{document}